\documentclass[]{aastex701}
\usepackage[utf8]{inputenc}
\usepackage[titletoc, title]{appendix}
\usepackage{CJKutf8}
\usepackage{blindtext}
\usepackage{amsmath}
\usepackage{mathtools}

\def \MIDAS {Michigan Institute for Data and AI in Society, University of Michigan, Ann Arbor, MI 48109, USA}
\def \Physics {Department of Physics, University of Michigan, Ann Arbor, MI 48109, USA}
\def \Astronomy {Department of Astronomy, University of Michigan, Ann Arbor, MI 48109, USA}
\def \CfA {Center for Astrophysics | Harvard \& Smithsonian, 60 Garden Street, Cambridge, MA 02138, USA}
\def \CWRU{Department of Physics, Case Western Reserve University, Cleveland, OH 44016, USA}

\graphicspath{{figures/}}

\begin{document}
\shorttitle{\texttt{heliostack}: A Novel Approach to Minor Planet Discovery}

\title{\texttt{heliostack}: A Novel Approach to Minor Planet Discovery}

\correspondingauthor{Kevin J. Napier}
\author[0000-0003-4827-5049]{Kevin~J.~Napier}
\affiliation{\CfA}
\affiliation{\Physics}
\affiliation{\MIDAS}
\email[show]{kevin.napier@cfa.harvard.edu}

\author[0000-0002-1139-4880]{Matthew~J.~Holman}
\affiliation{\CfA}
\email{mholman@cfa.harvard.edu}

\author[0000-0001-7737-6784]{Hsing~Wen~Lin (\begin{CJK*}{UTF8}{gbsn}
林省文\end{CJK*})}
\affiliation{\Physics}
\email{hsingwel@umich.edu}

\author[0000-0001-6942-2736]{David W. Gerdes}
\affiliation{\CWRU}
\affiliation{\Physics}
\affiliation{\Astronomy}
\email{gerdes@umich.edu}

\author[0000-0003-0403-0891]{Thomas~R.~Ruch}
\affiliation{\Physics}
\email{trruch@umich.edu}

\begin{abstract}
The study of faint solar system objects is a promising avenue for understanding the origin and evolution of planetary systems. However, such objects are difficult to detect in conventional surveys. Here we introduce \texttt{heliostack}, an algorithm for nonlinear shift-and-stack searches for solar system objects, which enables us to combine images taken over longer time spans than was previously possible. Applying this algorithm to a number of existing archival and forthcoming surveys will allow us to maximize their potential for discovering faint solar system objects. In this work, we apply \texttt{heliostack} to archival Hubble Space Telescope (HST) data, completing an exhaustive search for Cold Classical Kuiper Belt Objects in a set of HST images taken over a 15-day time span in 2003. We successfully recover both of the known sub-threshold objects in the data, and add two new discoveries. These two new objects are the first to ever be discovered in stacks of images taken over a time span longer than about one day.
\end{abstract}

\accepted{15 April 2026}
\submitjournal{Planetary Science Journal}

\section{Introduction}
\label{sec:intro} 

Our solar system's small body populations offer a unique perspective for studying planetary systems. Today the census of minor planets stands at more than one million, which when taken in aggregate, reveal rich and delicate dynamical structures that have been sculpted over billions of years by the gravity of the planets. By studying the dynamical and physical properties of our solar system's populations of near-Earth objects, asteroids, Jupiter Trojans, centaurs, comets, Neptune Trojans, Kuiper Belt Objects, and beyond, we have gained deep insight about the processes driving the formation and evolution of planetary systems. Still, fundamental questions remain: Is there a massive planet still awaiting discovery in the distant solar system \citep{batygin2019}? Are there more populations of small bodies that have thus far eluded detection? To answer these questions we must develop techniques to discover fainter objects so that we can probe smaller and more distant populations. 

In the two centuries since scientists began working to detect and characterize solar system bodies, progressive improvements in telescope and detector technology have greatly enhanced our capabilities. However, we are reaching the physical limits of detectors and the practical engineering limits of telescopes, and since the objects we are searching for are moving, we cannot simply take longer exposures to increase image depth. One way that we can continue to make progress without improving hardware, is by developing software and algorithms that can better utilize our data. 

There is a standard procedure for detecting moving objects fainter than the single-image detection threshold, called shift-and-stack, in which one stacks images along the trajectory of some putative object \citep{tyson1992, gladman1998, allen2002, bernstein2004, holman2004, smotherman2021, napier2024b}. When the signal from the object is combined in a sufficient number of images, the object becomes detectable. The problem, however, is that we do not usually have \textit{a priori} knowledge of the trajectories of the real objects in our images. In those cases, we must search through \textit{all possible} orbits of interest. Historically, shift-and-stack searches for solar system objects have not explored a full six-dimensional orbital parameter space, but have instead remained limited to a four-dimensional regime where an object's apparent motion remains linear in pixel space. In this regime, it is simple to determine which trajectories need to be checked for an exhaustive search (see \citealt{napier2024b}, for example). However, the requirement for linear trajectories limits stacks to data taken over the span of a few hours. When the duration of the exposure sequence gets any longer, the apparent motion of the objects becomes nonlinear in pixel space, increasing the complexity of the problem. Because of the increased complexity over longer time spans, the shift-and-stack technique has thus far been restricted to dedicated shift-and-stack surveys, which typically take of order 100 images of the same patch of sky in a given night \citep{DEEPI, DEEPII}


To date, only \citet{bernstein2004} has completed a systematic shift-and-stack search for solar system objects in images that were taken over a time span longer than a few hours. There has been recent work in the area of nonlinear digital tracking (see, e.g.~\citealt{Golovich2025}), but so far no systematic searches have been completed, and no new objects have been discovered. In the two decades since \citet{bernstein2004}, with the emergence of GPUs and incredible improvements in computing capacity on CPU clusters, why haven't any other shift-and-stack searches over long time baselines been completed? The answer may simply be that the problem is difficult from both theoretical and computational perspectives. Just enumerating the trajectories to be searched requires a careful choice of parameters, and even with an efficient parameterization the number of trajectories to search can quickly become unmanageable.

Despite its immense difficulties, the problem is worth solving. If we can manage to combine images taken days, months, or even years apart, we will be able to reach the depth of a shift-and-stack search using data from surveys with sparser cadences such as the Dark Energy Survey (DES), or the Legacy Survey of Space and Time (LSST). If we can gain one or two magnitudes of sensitivity over such large survey areas, we will see a substantial increase in the yield of solar system objects, with especially important gains for populations of objects that are small or very distant \citep{DEEPII, Juric2019}. Furthermore, if we can manage to search for solar system objects in sparser or more irregular image cadences, we can develop much more flexible survey plans, making it easier for the community of solar system observers to obtain time on the most oversubscribed telescopes such as the James Webb Space Telescope (JWST) or the Nancy Grace Roman Space Telescope (Roman).

In this paper we introduce \texttt{heliostack}, an algorithm designed to enable nonlinear shift-and-stack searches for solar system objects. One of the key insights that enables \texttt{heliostack} to efficiently search through the space of nonlinear trajectories is inspired by the eponymous \texttt{HelioLinC} algorithm \citep{holman2018}. Given detections of some solar system object, guesses of an object's unmeasured orbital parameters can be used to fully define an orbit for each detection. If the guess was correct (or at least close enough), those orbits will cluster on the sky when propagated to a common epoch. In retrospect, the idea is obvious---no object can be in two places at once! We take this idea a step further in \texttt{heliostack}, applying the idea not just to catalogs of tracklets, but rather treating \textit{each pixel} in each image as a ``detection'', and using some set of asserted orbital parameters to synchronize our pixels to a common epoch.

We begin by describing the \texttt{heliostack} algorithm in Section \ref{sec:algorithm}. In Section \ref{sec:hst}, we demonstrate the algorithm by searching for Kuiper Belt Objects (KBOs) in the Hubble Space Telescope (HST) data from \citet{bernstein2004}. Finally, in Section \ref{sec:discussion}, we identify some immediate applications of the \texttt{heliostack} algorithm, and speculate as to how it may be useful in the future, as both data volume and computing capacity continue to grow.

\section{The Algorithm}

\label{sec:algorithm}

We operate under the hypothesis that we have some images that contain an object that is too faint to be confidently detected in individual exposures. Our goal is to either find this object or, in failing to do so, place some upper limit on how bright it may be. In order to do this, we need to have prior knowledge of the object's location in each of our images, but fortunately we do \textit{not} need to have prior knowledge of its brightness. While objects' brightnesses can be variable in complicated ways, their trajectories are usually deterministic from a given initial condition, and are therefore simple to describe.\footnote{We consider only Keplerian orbits in the present treatment.} In particular, the motion of solar system objects can be described by some appropriate set of six orbital parameters at a given time. But how can we choose our parameters to completely, and non-redundantly, trace out the paths that any possible moving object may take on the sky as seen by some moving observatory? The problem is complicated, and becomes even more complicated if one considers observations from multiple observatories. We can simplify the problem by using an approach inspired by \texttt{HelioLinC}. 

Let us begin with the idea of generating a blank canvas of the \textit{barycentric} sky at some reference epoch, $t_{\mathrm{ref}}$. Locations on this canvas can be described by unit vectors with an origin at the barycenter. We will use this canvas to accumulate signal from our images, so it should be made up of discrete bins (preferably pixels of approximately the same angular size as the pixels in the images that are being combined). By stacking images onto this canvas, we can decouple any topocentric observer information. But how do we determine where on our canvas each pixel from each image should go? 

First, we can recognize that a pixel in an image is associated with a topocentric sky position $\{ \alpha, \delta \}$, a time $t$, a flux, a flux uncertainty, and a vector describing the observer's position relative to some inertial frame, $\vec{x}_\mathrm{obs}$. The topocentric sky position of the pixel can be expressed as a Cartesian unit vector,
\begin{equation}
    \mathbf{\hat{\rho}} = \langle \cos{\delta}\cos{\alpha}, \cos{\delta}\sin{\alpha}, \sin{\delta} \rangle.
    \label{eq:rhohat}
\end{equation}
The barycentric position vector to the object at the time when it reflected the observed photons can be expressed as
\begin{equation}
    \vec{r}\left( t - \frac{\rho}{c}\right) = \vec{x}_\mathrm{obs} + \rho \hat{\rho}.
    \label{eq:light-time}
\end{equation}
where $\rho$ is the topocentric distance to the object, and $\rho / c$ is the light time. 

The direction of the object's barycentric unit vector \textit{at the reference epoch} dictates where it should be registered to on the canvas. In order to properly register all pixels to the canvas, then, we need to know $\vec{r}(t)$, which requires knowing all six orbital parameters. Notice that two of the required six pieces of information can come directly from the image itself, $\alpha$, and $\delta$. In order to fully specify the orbit, then, we need to assert four more orbital parameters. 

The first three orbital parameters we choose to assert are the barycentric distance $r$, the line of sight velocity $v_r$, and the tangential velocity, $v_\Omega$, which are components of a basis that is described in detail in \citealt{napier2024}. Under the assumption of two-body motion, the universal anomaly, $\chi(t)$, can be calculated at any time $t$ as long as we know $r$ and $v_r$ at a reference epoch $t_0$, the specific energy, $\varepsilon$, and the time interval, $t - t_0$ \citep{danby}. The expression for the position vector as a function of time can be written as
\begin{equation}
    \vec{r}(t) = \vec{r}_0 f + \vec{v}_0g
    \label{eq:fg}
\end{equation}
where
\begin{equation}
    f(t) = 1 - \frac{\chi(t)^2}{r_0} C(z)
\end{equation}
and
\begin{equation}
    g(t) = (t - t_0) - \frac{\chi(t)^3}{\sqrt{\mu}} S(z).
\end{equation}
Here $z = \chi(t)^2/a = -2\varepsilon\chi(t)^2/\mu$, and $C$ and $S$ are the Stumpff functions \citep{danby}. We can take the vector dot product of both sides of Equation \ref{eq:fg} to yield an equation for the scalar barycentric distance, $r$, at any given time:
\begin{equation}
    r(t)^2 = r_0^2 f(t)^2 + v_0^2g(t)^2 + 2f(t)g(t) r_0 v_{r,0}.
\end{equation}

The fourth parameter to assert is the orbital inclination, $i$, as well as a value called $\kappa \in \{ +1, -1 \}$, which breaks a degeneracy between ascending and descending orbits at the epoch of the elements (see \citealt{napier2024}). Our asserted $\{ r, v_r, v_\Omega, i\}$ map onto the orbital invariants. An orbit's specific energy, $\varepsilon$, specific angular momentum, $h$, and the $z$-component of the angular momentum, $h_z$, can be expressed as
\begin{equation}
    \varepsilon = \frac{v_r^2 + v_\Omega^2}{2} - \frac{\mu}{r}, \qquad h = r v_\Omega, \qquad \text{and} \qquad h_z = r v_\Omega \cos{i}.
    \label{eq:invariants}
\end{equation}
The mapping of our asserted orbital parameters onto the Keplerian invariants is an extremely useful property because it means that we can simultaneously solve for the orbits of \textit{all} pixels from \textit{all} images, making the problem easy to parallelize.

There is at most one orbit that has the observed or known quantities $\{ \alpha, \delta, \vec{x}_\mathrm{obs}, t \}$, and the asserted quantities $\{ r, v_r, v_\Omega, i, \kappa, t_{\mathrm{ref}}\}$, which we can solve for as follows.\footnote{If a pixel's latitude exceeds its putative inclination, no object with the chosen orbital parameters could possibly be on that pixel, and so we can simply ignore it.}  First we can take the vector dot product of Equation (\ref{eq:light-time}) with itself to put it into scalar form:
\begin{equation}
    \left[r\left( t - \frac{\rho}{c}\right)\right]^2 = |\vec{x}_\mathrm{obs}|^2 + \rho^2 + 2\rho (\hat{\rho} \cdot \vec{x}_\mathrm{obs}).
    \label{eq:light-time-scalar}
\end{equation}
It is possible to find a value of $\rho$ that solves Equation (\ref{eq:light-time-scalar}) because as previously established, $r(t)$ is uniquely determined at all times.\footnote{Note that in the case of a Keplerian orbit, $r(t)$ is completely decoupled from the orbit's orientation in space. This property enables the pre-calculation of $r(t)$ at all times for all possible sky locations, thus eliminating a lot of redundant computation.} Plugging our solution for $\rho$ back into Equation (\ref{eq:light-time}), we can get the position vector to the orbit at the epoch when the object reflected the observed photons. To completely specify the orbit, we also need to find its velocity vector at the light-corrected epoch. We can do so by using the approach described by \citet{napier2024}, who show that we can express the state vector as 
\begin{align}
    \vec{r} &= r \hat{r}
    \label{eq:position} \\
    \vec{v} &= v_r \hat{r} + v_{\Omega} \left( \cos{\psi} \hat{A} + \sin{\psi} \hat{D}\right)
    \label{eq:velocity}
\end{align}
where we have now introduced $\psi$, which describes the direction that $v_{\Omega}$ points in the plane defined by $\hat{A}$ and $\hat{D}$, which are given by
\begin{align}
    \hat{A} &= \langle -\sin{\varphi}, \cos{\varphi}, 0 \rangle
    \\
    \hat{D} &= \langle -\sin{\theta}\cos{\varphi}, -\sin{\theta}\sin{\varphi}, \cos{\theta} \rangle.
    \label{eq:AD}
\end{align}
We already have $\{r, \varphi, \theta\}$ at the light-corrected epoch from Equation (\ref{eq:light-time}), so the remaining undetermined parameters are $v_r$, $v_\Omega$, and $\psi$. Since we have specified the orbital invariants, we can solve for those remaining parameters as follows:
\begin{align}
    v_\Omega &= \frac{h}{r} \\
    \epsilon &= \frac{v^2}{2} - \frac{\mu}{r} \implies v^2 = 2\left(\epsilon + \frac{\mu}{r}\right) \\
    v_r &= \sqrt{v^2 - v_\Omega^2} \\
    \cos\psi &= \frac{\cos i}{\cos{\theta}} \\
    \sin\psi &= \kappa \sqrt{1 - \cos^2\psi}. 
\end{align}
Now the orbit's position and velocity at the light corrected epoch are fully specified, so it can be propagated to the reference epoch of the initial condition and we can accumulate the signal onto the canvas accordingly. Each pixel in the canvas corresponds to an orbit which has orbital parameters given by the pixel's $\varphi$ and $\theta$, and the initial condition's $r$, $v_r$, $v_\Omega$, $\psi$, and $t_\mathrm{ref}$. Any pixel with sufficiently high significance can be thought of as a candidate moving object, which can later be investigated more carefully.


\section{Demonstration on Archival HST Data}
\label{sec:hst}

As a proof of concept for the \texttt{heliostack} algorithm, we have reprocessed the data from \citet{bernstein2004} in a targeted search for Cold Classical Kuiper Belt Objects (CCKBOs). The images from this dataset were obtained in 2003 during 125 HST orbits over a time span of about 15 days, and cover about 0.02 square degrees with six dithered, nearly-contiguous ACS WFC fields of view near the solar system's invariable plane. Each of the survey's pointings was imaged for a total integration time of about 38,000 seconds, broken into 95 exposures of approximately 400 seconds each. These images were strategically taken near quadrature, so that the TNOs in the field would be near their semi-annual parallax turnaround point at the midpoint of the survey, preventing them from drifting out of the survey's extremely small field of view. More detailed information about the images can be found in \citet{bernstein2004}. We used the MAST archive to acquire the \texttt{.flc} images of this field, which were processed by HST's \texttt{CALACS} pipeline, and thus were calibrated, CTE-corrected, and given internally consistent astrometric solutions.

When \citet{bernstein2004} processed these data more than 20 years ago, it was computationally prohibitive to stack images taken over a span of more than 24 hours. To compensate for this limitation, they stacked chunks of the data, and then linked the stacked detections across multiple epochs. In this work, since it is now computationally feasible, we simply stack the entire dataset.

\subsection{Image Preprocessing}

We had to take a few simple steps to prepare our images to be searched. At each step, we treated both our images their corresponding variance so that we could properly weight our images in the stacks. We began by injecting a synthetic population of CCKBOs for later use in quantifying our detection efficiency. It is best for this step to occur before any further processing, so that the synthetic sources undergo the same processing as real sources to the extent possible. Our synthetic population had apparent magnitudes between 28.4 and 29.7 at the survey's midpoint, and single-component sinusoidal light curves with amplitudes as large as 1 magnitude, and periods between 4 and 36 hours. We implanted the flux using an empirical PSF model \citep{epsf} with appropriate trailing.

The rest of the steps can be enumerated as follows:
\begin{enumerate}
    \item Convert all images to units of electrons per second.
    \item Fit and subtract a spatially-varying background from each image.
    \item Mask regions with 500 or more contiguous pixels brighter than $5\sigma$. 
    \item Resample each image onto a rectilinear grid with a uniform pixel scale of $0.05''$, using Lanczos-4 interpolation.
    \item Generate a template for each CCD by aligning the images and calculating a robust Huber mean \citep{Huber1964}.\footnote{We found that the robust mean was successful at preventing transients (including both cosmic rays and real moving solar system objects) from appearing in the template; our templates caused no measurable self-subtraction of our implanted synthetic objects.}
    \item Create difference images by subtracting the corresponding template from each image.
    \item Extract all sources brighter than 3.5$\sigma$ from each image. These catalogs could be used for a search for bright objects, but we do not do a bright search in this analysis.
    \item Flag and mask all pixels brighter than 4$\sigma$, as well as a 2-pixel buffer. This step is important for avoiding false positives in the stacks.
\end{enumerate}
Since HST's PSF and seeing are so stable, the image subtraction worked well without any PSF matching or rescaling. After masking bright transients (i.e.~cosmic rays, asteroids, TNOs), we find that each of our images was well approximated as Gaussian noise without correlations. This was a critical detail for avoiding false positives in the stack.

Finally, we generated two new kinds of images, defined as follows:
\begin{equation}
    \xi = \left(\frac{D}{V}\right) \star P
\end{equation}
and
\begin{equation}
    \zeta = \left(\frac{1}{V}\right) \star P^2,
\end{equation}
where $D$ is the difference image, $V$ is its variance, $P$ is the image's PSF, and $\star$ represents a cross correlation operation. These pre-convolved $\xi$ and $\zeta$ images are optimal for point-source detection, and make it trivial to calculate a weighted average with a PSF-matched filter when stacking \citep{bernstein2004, Lang2026}. Notice that $\zeta$ is, by definition, just the variance of $\xi$, so the significance of any pixel is given by $\nu = \xi / \zeta^{0.5}$. One drawback to treating the images this way is that since we only convolve the images with their PSF one time, they are not quite optimal for the detection of trailed sources. This limitation is not a big issue though, because after the initial identification of candidates, it is easy to investigate them more carefully (see Section \ref{sec:tuneup}).

\subsection{Grid of Initial Conditions}

The final bit of preparation was to choose which orbits to search. Since the space of orbital parameters is continuous there are, in principle, infinitely many possibilities. In practice, however, the search space consists of a finite set of initial conditions that guarantees tracking errors to be smaller than some tolerance $\varepsilon$ for all points in the parameter space of interest. That is, no trajectory of interest will deviate from a searched trajectory by more than $\varepsilon$. For this search, we chose $\varepsilon = 0.05 ''$, comparable to the PSF width and pixel scale of ACS WFC. Tracking errors at this level can, in principle, degrade the depth of our search by up to 0.1-0.2 magnitudes, but stacking at a finer resolution, for instance $\varepsilon = 0.03 ''$ as done by \citet{bernstein2004}, would represent a fourfold increase in computing resources which we consider to be unnecessary for this proof of concept.

For this demonstration, we generated a grid of initial conditions by selecting values of $r$, $v_\Omega \cos{\psi}$, and $v_\Omega \sin\psi$ that yield tracking errors smaller than $0.05''$ over the 15-day time baseline of our data \citep{Holman2019}. Over such short arcs, the radial velocity of the CCKBOs does not make any measurable difference to their sky motion, so we do not lose any sensitivity by using $v_r = 0$ for each initial condition. The problem becomes more complicated for longer arcs where deviations in $v_r$ become detectable, but it is still solvable (see \citealt{ruch2026} for one approach). We pare our initial conditions to the parameter space of the CCKBOs with bounds $a \in [42.4, 47.4]$ au, $e \in [0, 0.2]$, and $i_{\rm{ecliptic}} \in [0, 5]$ degrees. We also restrict $r \in [38, 51]$ au, as this range includes more than 95\% of the CCKBOs \citep{Kavelaars2021}. These restrictions result in a total of 1,148,817 initial conditions to check. 
Checking the full Kuiper belt beyond 38 au, with inclinations less than $60^{\circ}$, would require a factor of 300 more initial conditions (though it is likely that the grid could be made smaller by eliminating some redundant trajectories).


\subsection{Stacking and Candidate Identification}
\begin{figure}[ht]
    \centering
    \includegraphics[width=1.0\linewidth]{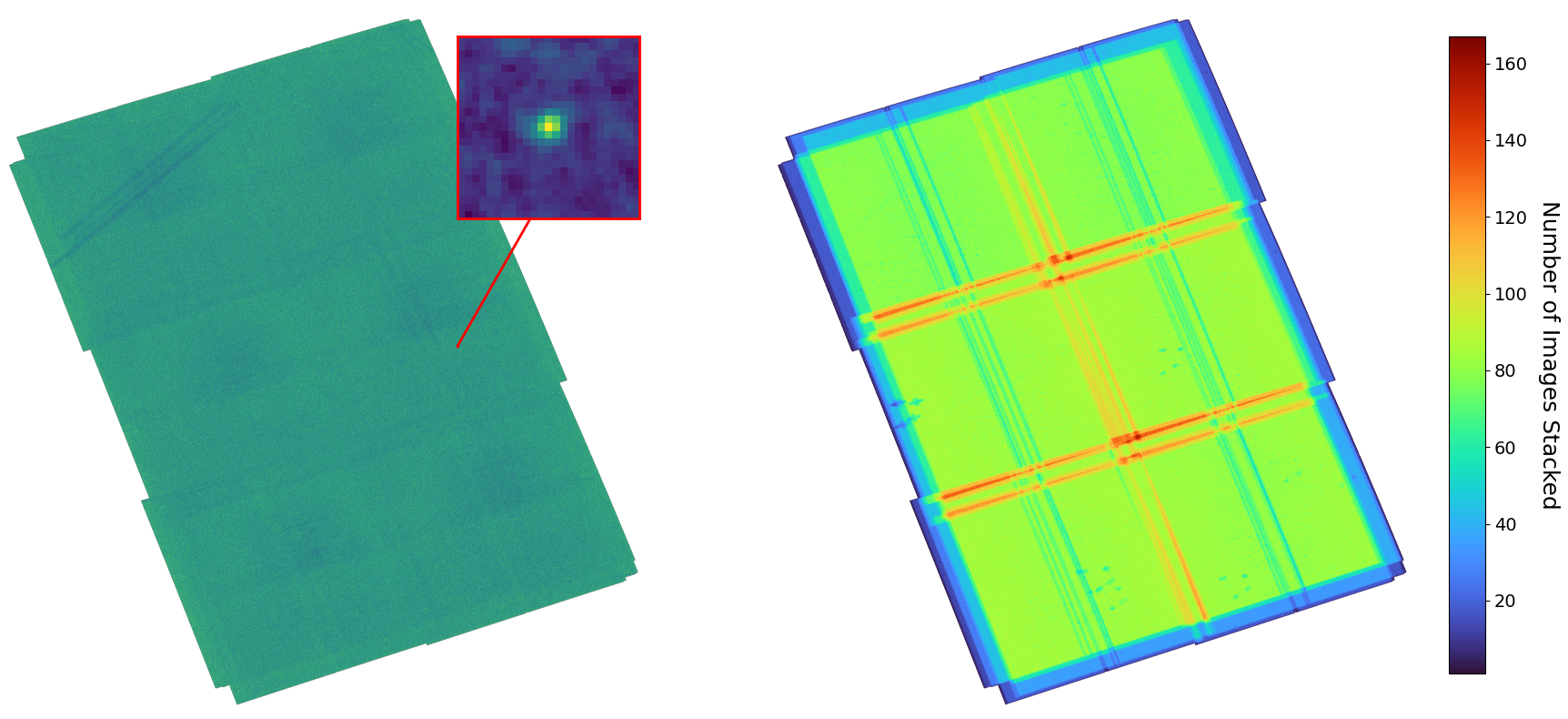}
    \caption{(Left): The canvas (approximately 14,000 by 13,000 pixels) after stacking all 1118 images for some initial condition. The inset plot shows the lone peak above 7$\sigma$ in the stack, which turns out to be a $21\sigma$ detection of 2003 $\mathrm{BH}_{91}$. Since the asserted parameters are not quite correct, the detection is somewhat smeared. (Right): A heatmap of the number of images that contributed to each pixel in the significance image.}
    \label{fig:stack}
\end{figure}

With our images pre-processed and a list of orbits to check, we proceeded with the search. For each orbit, we began by generating a blank canvas with a linear WCS of barycentric $(\varphi, \theta)$ at a reference epoch of the survey's midpoint. \footnote{The tangent plane is arbitrarily defined with respect to the lower left corner of the canvas in each stack.} We then synchronized each image to our canvas using the transformation described in Section \ref{sec:algorithm}. In principle we could have done this synchronization pixel-by-pixel. However, since our images were reprojected to have a linear WCS in the preprocessing stage, we were able to instead calculate the transformation for each of the corners of an image, and use the transformed points to calculate the perspective transformation matrix to project the topocentric coordinates onto the barycentric canvas (see, e.g., \citealt{Hartley_Zisserman_2004}).\footnote{We find empirically that errors introduced by the perspective projection (as opposed to going pixel-by-pixel) remain smaller than 0.2 pixels over a full image, providing plenty of fidelity for initial candidate identification.} This approximation provides a significant speedup, especially on GPU, because it takes advantage of the fact that images are spatially indexed and contiguous in memory.  We then applied this perspective transformation to the image, and added the image to our canvas accordingly, keeping running sums of both $\xi$ and $\zeta$. Since this stage of the search is very approximate, we used nearest neighbor interpolation to make the stacks as fast as possible. More complicated interpolation schemes would also work without much of a performance penalty, but the exact resampling scheme does not matter much at this stage in the process. After all of the images were synchronized to the canvas for a given orbit, we calculated a significance image, $\nu = \xi / \zeta^{0.5}$, ran a peak finding routine to find all local maxima on the canvas, and took note of all peaks with $\nu \geq 7.0$. For an example of the significance image, see the left-hand panel of Figure \ref{fig:stack}.

Searching the 1118 images with 1,148,817 initial conditions yielded a total of 109,383 candidate objects, each associated with a complete orbit specification, $\{\varphi, \theta, r, v_r, v_\Omega, \psi\}$ at $t_{\mathrm{ref}}$, as well as a flux and a flux uncertainty. Many of these candidates turned out to be redundant detections of the same object, but found with slightly different significances and with different asserted parameters. To eliminate such repeated detections, we identified objects whose trajectories coincided to within 0.2'' in both RA and Dec in at least 30 images, and retained the candidate with the highest $\nu$ from each group of associated trajectories. This step resulted in a reduction to 533 total candidates with $\nu \geq 7.0$. Still, a few of the trajectories were redundant detections of the same object, but found along different segments of the object's trajectory; these duplicates are resolved after the orbit tuneup procedure described in Section \ref{sec:tuneup}. Of the 533 candidates, 165 would later be identified as implanted sources, meaning that there were 368 sources remaining that were not associated with an implanted synthetic source, which is roughly consistent with the expected number of sources with $\nu > 7$, assuming Gaussian noise and $\sim$$2\times10^{14}$ trials (see the distribution of $\nu$ in the top panel of Figure \ref{fig:snr}). It is a testament to the quality of these data, and the importance of using clean difference images, that the candidate list was so manageably small without any machine learning or visual inspection for false positive rejection. At this point, each of our candidates was still just associated with a \textit{discrete} orbit from our list of initial conditions, and needed to be tuned up in a continuous parameter space.

\subsection{Candidate Tuneup}
\label{sec:tuneup}
\begin{figure}
    \centering
    \includegraphics[width=0.9\linewidth]{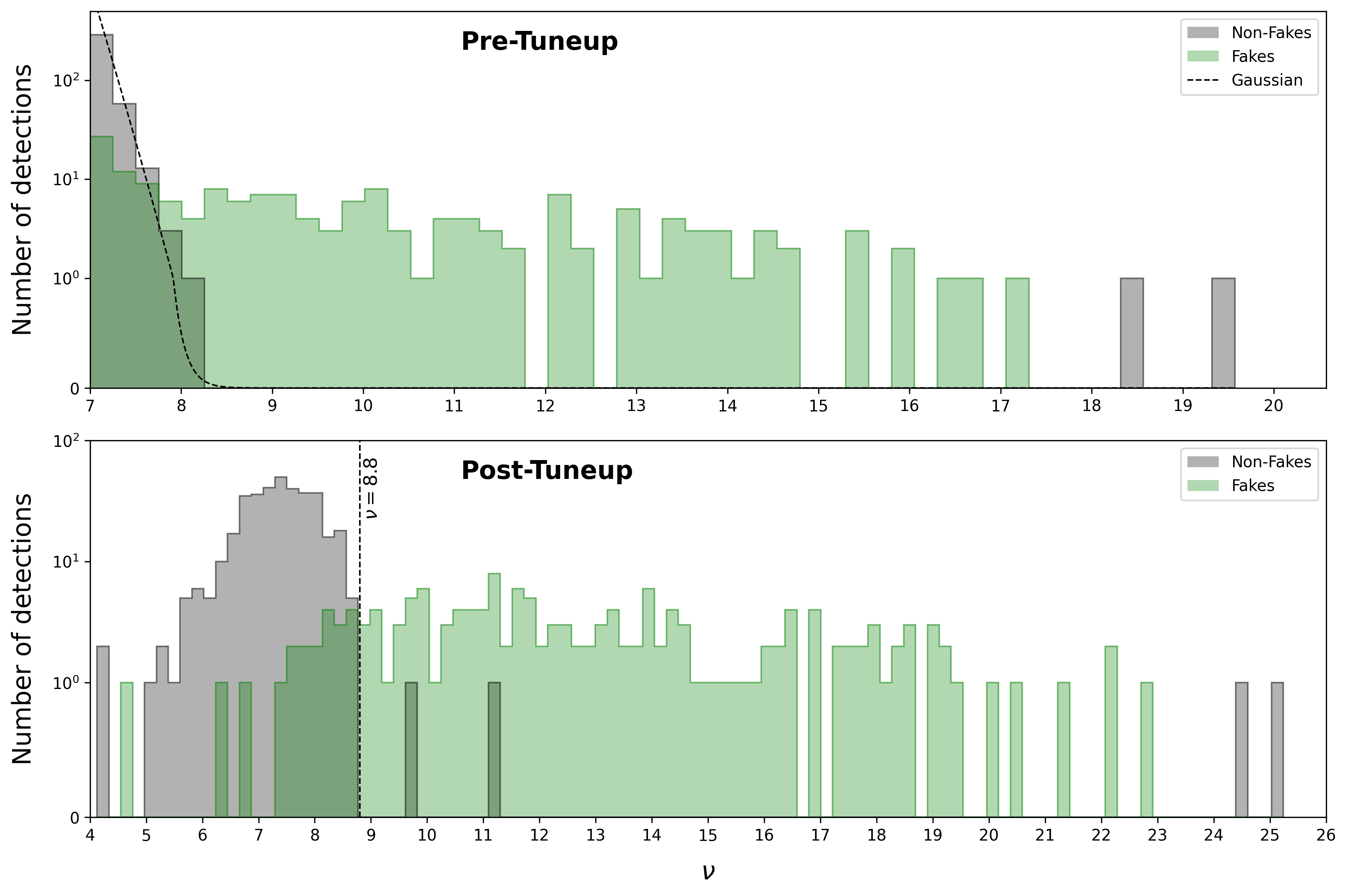}
    \caption{(Top) Distribution of the signal-to-noise ratio of all of the unique candidate trajectories resulting from the \texttt{heliostack} search, with $\nu \geq 7$. The green histogram represents the implanted synthetic sources, and the gray histogram represents real candidate moving objects. The dashed black line shows the shape of a Gaussian distribution with mean 0 and width 1. It is clear that the distribution of $\nu$ is very nearly Gaussian. (Bottom) Distribution of the signal-to-noise ratio of all of the unique candidate trajectories after the tuneup step described in Section \ref{sec:tuneup}. Real candidates tend to increase in significance, while spurious candidates tend not to. There is a clear boundary at $\nu = 8.8$, beyond which there are probably no false positives.}
    \label{fig:snr}
\end{figure}

After finding our candidates in the discrete search, we optimized each source's significance by fitting a 6-parameter orbit directly on the pixels. For this tuneup procedure we used a different stacking approach, working directly with the non-convolved difference images. For each candidate, we chose a 6-parameter orbit given by $\{ \alpha, \delta, r, v_r, v_\Omega, i, \kappa \}$ at $t_{\mathrm{ref}}$, propagated the orbit to the epoch of each image, and used a trailed PSF model to obtain forced measurements of the flux and flux uncertainty of the putative source.\footnote{Using the topocentric observables helped to constrain the solutions, because the observer's position is fixed at the reference epoch, meaning that the solution for an observation's barycentric sky position $\varphi$, $\theta$ is coupled to its barycentric distance.} This process generated a light curve for each object, which we then used to calculate a sigma-clipped (3$\sigma$ with 10 iterations) mean flux and flux uncertainty. Taking the ratio of the mean flux to the flux uncertainty yields a significance, $\nu$, and we used $\nu^2 / 2$ (which is the Gaussian log likelihood) as an optimization metric for a downhill simplex optimizer. The result of this optimization procedure is shown in the bottom panel of Figure \ref{fig:snr}. In general, real candidates tend to increase in significance, while spurious candidates tend not to. There is a clear boundary at $\nu = 8.8$, beyond which there are likely no false positives. Note that one major difference between the top and bottom panels is that the $\nu$ values in the tuneup procedure were calculated using a robust (sigma-clipped) mean, while the initial candidates' significances were derived from inverse variance-weighted means, with no outlier rejection; This is the reason why some candidates have $\nu < 7$ after tuneup.

\subsection{Results}

\begin{figure}[ht]
    \centering
    \includegraphics[width=0.9\linewidth]{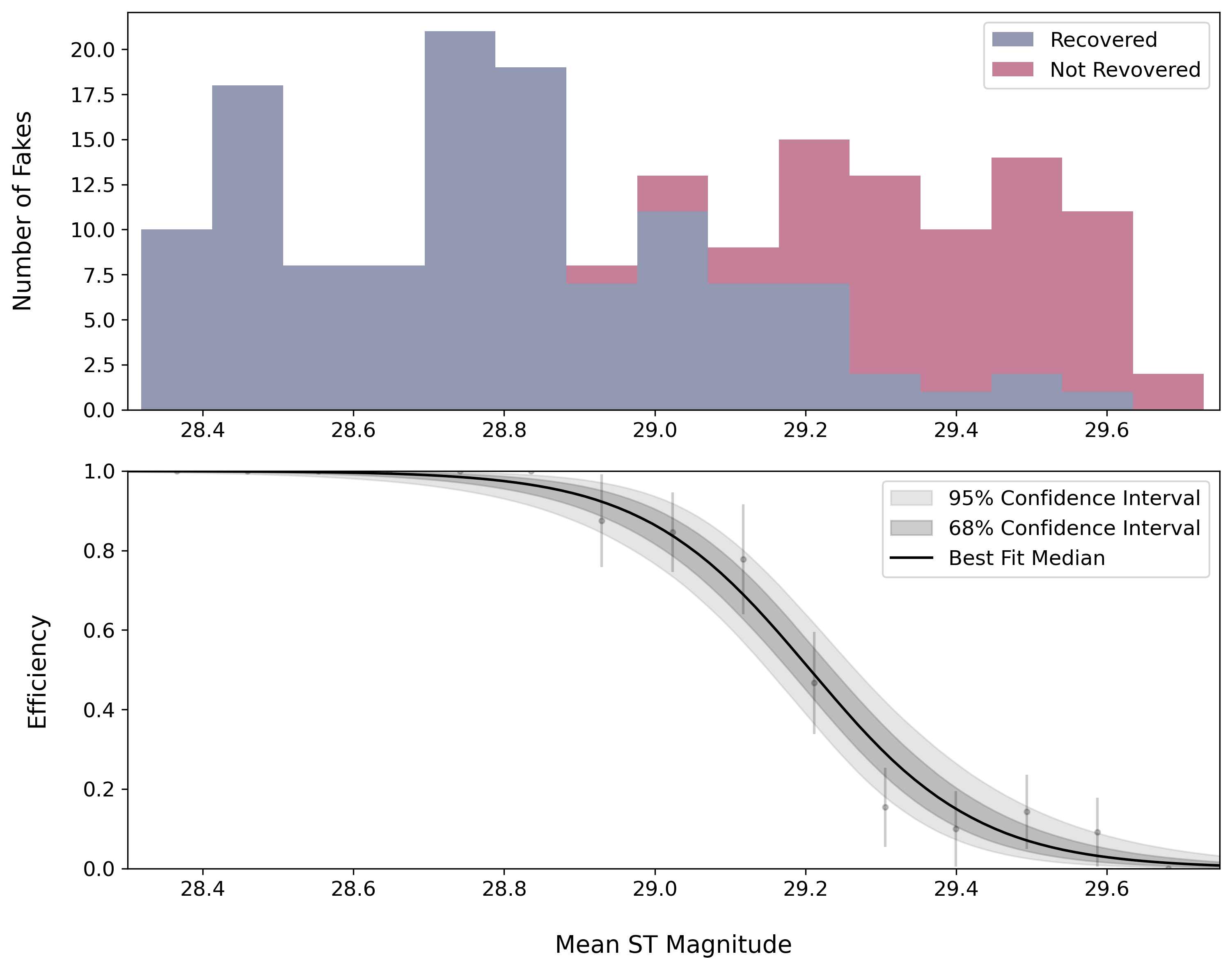}
    \caption{(Top): Histogram of the mean apparent magnitude of the implanted synthetic sources, with an accounting of whether or not the sources were recovered. (Bottom): Best fit of our efficiency to Equation (\ref{eq:efficiency}), including 68\% and 95\% confidence intervals. The best-fit parameters are $\eta_0 = 1.0$, $m_{50} = 29.21 \pm 0.03$, and $\sigma = 0.11 \pm 0.02$.} 
    \label{fig:efficiency}
\end{figure}

After tuneup, we retained all candidates with $\nu \geq 8.8$, which is well-motivated by Figure \ref{fig:snr}. Our final list contained 145 candidates, 141 of which turned out to be implanted sources. Our efficiency, shown in Figure \ref{fig:efficiency}, is reasonably well-described by a function of the form
\begin{equation}
    \eta\left(m\right) = \frac{\eta_0}{1 + \exp\left(\frac{m - m_{50}}{\sigma}\right)}
    \label{eq:efficiency}
\end{equation}
with $\eta_0 = 1.0$, $m_{50} = 29.21 \pm 0.03$, and $\sigma = 0.11 \pm 0.02$. Note that we used a maximum likelihood approach to fit our efficiency, rather than binned data. One notable feature of our efficiency function is a rather long tail, which can be attributed to increased sensitivity in the regions where chips overlap in the stack (see the right-hand panel in Figure \ref{fig:stack}).

\begin{figure}[ht]
    \centering
        \includegraphics[width=0.2\linewidth]{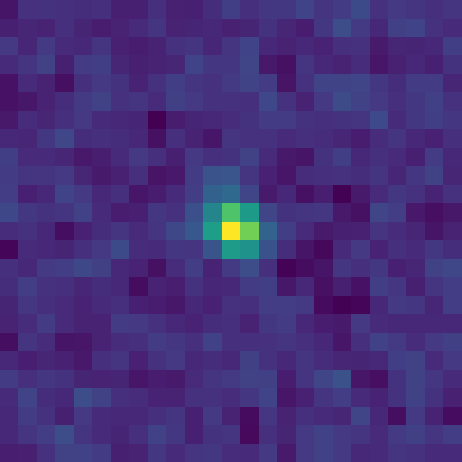}
        \hspace{0.5cm}
        \includegraphics[width=0.2\linewidth]{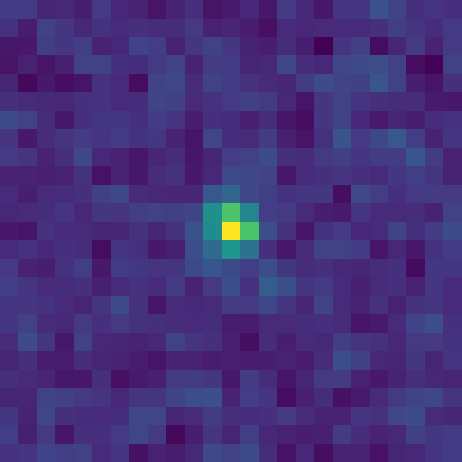}
        \hspace{0.5cm}
        \includegraphics[width=0.2\linewidth]{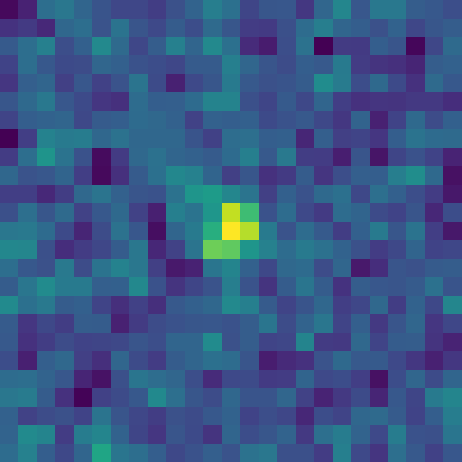}
        \hspace{0.5cm}
        \includegraphics[width=0.2\linewidth]{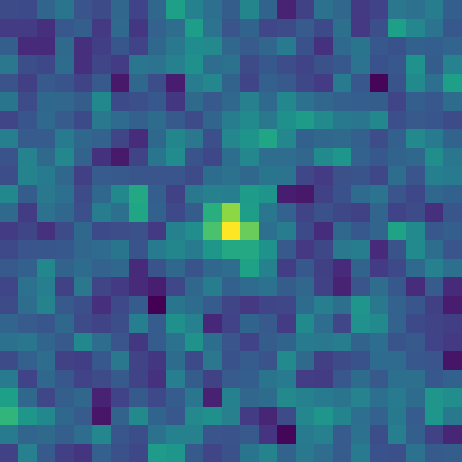}
    \caption{From left to right, stacked cutouts of 2003 $\mathrm{BF}_{91}$, 2003 $\mathrm{BH}_{91}$, 2003 $\mathrm{ABCD}$, and 2003 $\mathrm{WXYZ}$.}
    \label{fig:cutouts}
\end{figure}

\begin{figure}
    \centering
    \includegraphics[width=0.95\linewidth]{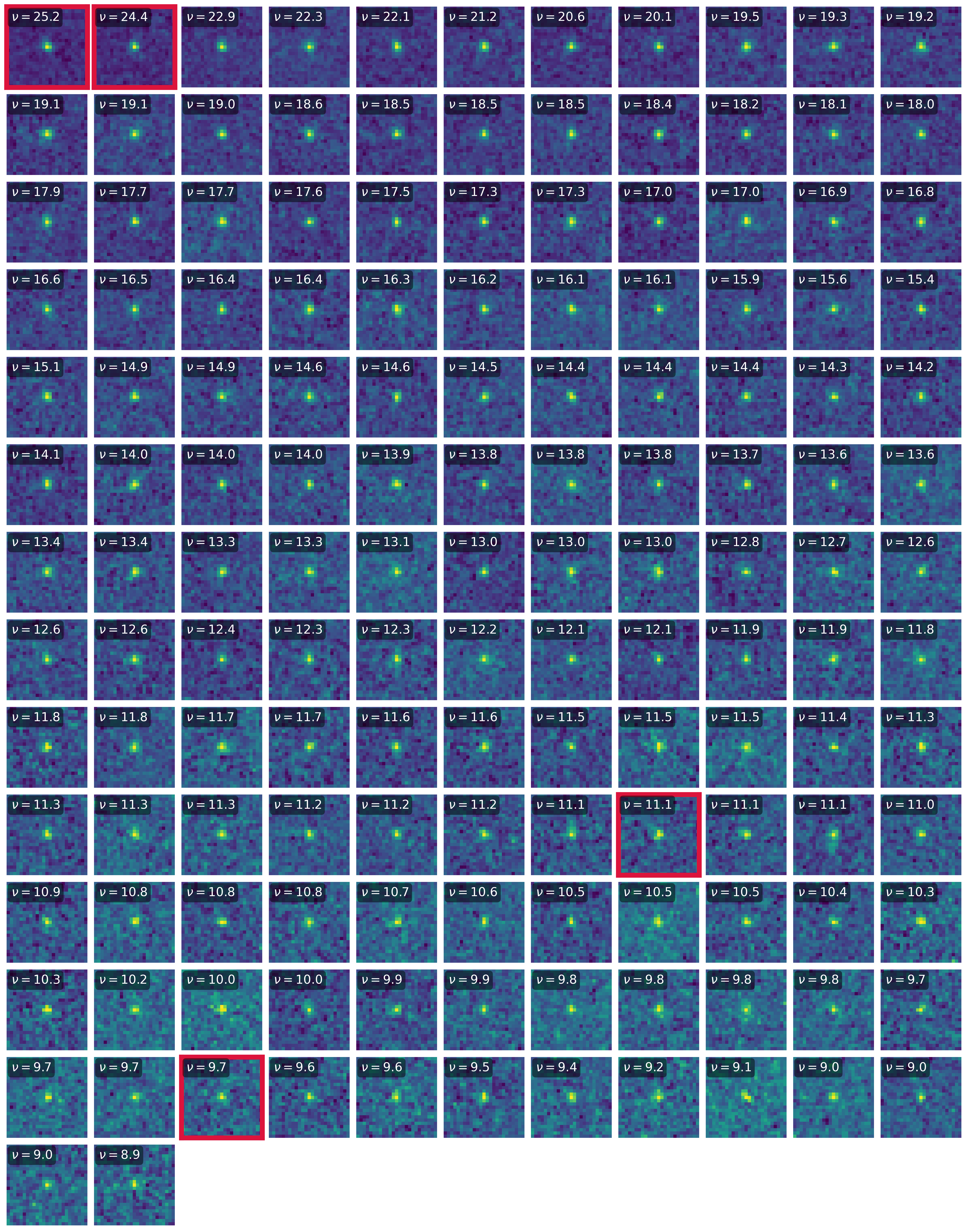}
    \caption{Cutouts of each of the recovered sources. Objects that are not implanted synthetic sources are highlighted in red.}
    \label{fig:fakes}
\end{figure}

The remaining four candidates are, as far as we can tell, real objects. Two of the objects, 2003 $\mathrm{BF}_{91}$ and 2003 $\mathrm{BH}_{91}$, were previously discovered by \citet{bernstein2004}. The other two objects, which we call 2003 ABCD and 2003 WXYZ as they do not yet have Minor Planet Center designations, are newly discovered in this search; they were likely missed in the search by \citet{bernstein2004} because they were simply too faint to be found without stacking all of the survey's images simultaneously. We show cutouts of the objects in Figure \ref{fig:cutouts}, and list their photometric and orbital properties in Table \ref{tab:object_parameters}. The cutouts of these objects are indistinguishable from the cutouts of the recovered synthetic sources, shown in Figure \ref{fig:fakes}. The uncertainties in the orbital parameters come from an MCMC simulation in which we allowed all six orbital parameters to vary, with uniform priors. Note that we did not bother searching for the two brighter objects in this dataset, 2000 $\mathrm{FV}_{53}$ and 2003 $\mathrm{BG}_{91}$, as we consider finding objects that can be seen in individual exposures to be a solved problem for data with this survey's cadence.

\begin{table}[ht]
\caption{Photometric and orbital parameters for detected objects.}
\centering
\label{tab:object_parameters}
\renewcommand{\arraystretch}{1.3}
\setlength{\tabcolsep}{10pt}
\begin{tabular}{lcrcccc}
\toprule
Object ID & Flux ($e^-/s$) & $\nu$ & total integration (s) & ST Magnitude & $r_{\mathrm{bary}}$ (au) & $i_{\mathrm{ecliptic}}$ (deg) \\
\hline
2003 $\mathrm{BF}_{91}$ & $0.299 \pm 0.012$ & 25.2 & 27,622 & $27.99 \pm 0.04$ & $42.19^{+0.01}_{-0.01}$ & $1.50^{+0.001}_{-0.001}$ \\
\hline
2003 $\mathrm{BH}_{91}$ & $0.255 \pm 0.010$ & 24.4 & 33,576 & $28.16 \pm 0.04$ & $42.59^{+0.02}_{-0.02}$ & $2.00^{+0.014}_{-0.013}$ \\
\hline
2003 $\mathrm{ABCD}$ & $0.119 \pm 0.011$ & 11.1 & 29,875 & $28.99 \pm 0.10$ & $48.95^{+0.05}_{-0.05}$ & $2.00^{+0.025}_{-0.028}$ \\
\hline
2003 $\mathrm{WXYZ}$ & $0.100 \pm 0.010$ & 9.7 & 32,458 & $29.17 \pm 0.11$ & $42.73^{+0.06}_{-0.05}$ & $3.04^{+0.015}_{-0.023}$ \\
\hline
\end{tabular}
\label{tab:photometry}
\end{table}

While \citet{bernstein2004} reported a limiting magnitude of $m_{50} = 29.17$, similar to our limiting magnitude of $m_{50} = 29.21 \pm 0.03$, we believe that our search actually went significantly deeper, and that there is some discrepancy with the magnitudes reported. The first piece of evidence comes from the measured magnitudes of the objects detected by both analyses. We measure the mean magnitude of 2003 $\mathrm{BF}_{91}$ to be $27.98 \pm 0.04$, while \citet{bernstein2004} reported $28.15 \pm 0.04$. Similarly, we measure the mean magnitude of 2003 $\mathrm{BH}_{91}$ to be $28.16 \pm 0.04$, while \citet{bernstein2004} reported $28.38 \pm 0.05$. Second, we have confident detections of both 2003 $\mathrm{ABCD}$ and 2003 $\mathrm{WXYZ}$, while \citet{bernstein2004} did not find either object. 

One possible reason for the magnitude discrepancy is some difference with the zeropoints used in the analyses. An erratum from \citet{bernstein2004} \citep{b04-erratum} may support this idea of mismatched zeropoints, though it is difficult to know for certain. It is also worth noting that the images have undergone different processing by more recent versions of the \texttt{CALACS} pipeline since the search by \citet{bernstein2004}. Regardless of the cause for the discrepancy, it seems that our search went about 0.2 magnitudes deeper. Since our stacks had an effective integration time of 38,000 seconds, compared to 22,000 seconds for the search by \citet{bernstein2004}, the theoretical maximum increase in depth was $1.25 \log_{10}(38,000/22,000) \approx 0.3$ magnitudes. Given that we used a significance cutoff of $\nu = 8.8$, rather than $\nu = 8.2$ as done by \citet{bernstein2004}, we would expect a loss of $-2.5 \log_{10}(8.2/8.8) \approx 0.08$ magnitudes, so our gain of about $0.2$ magnitudes rather than $0.3$ magnitudes is more-or-less expected. 

Of the five KBOs in this dataset (excluding the targeted KBO 2000 $\mathrm{FV}_{53}$), only 2003 $\mathrm{BG}_{91}$ can be securely classified as a CCKBO. The arcs are simply too short for the orbits to be well-determined. The other objects, 2003 $\mathrm{BF}_{91}$, 2003 $\mathrm{BH}_{91}$, 2003 $\mathrm{ABCD}$, and 2003 $\mathrm{WXYZ}$, may be CCKBOs given their measured barycentric distances and orbital inclinations, but cannot be definitively classified because their orbital elements are not well-enough constrained by the 15-day arcs. For the MCMC samples of each object where their $a$ and $e$ values were consistent with CC orbits, we calculated the free inclinations as prescribed by \citealt{Huang2022}. We report 95\% confidence intervals for this experiment, along with the minimum allowed values of $a$ and $e$, in Table \ref{tab:free_inc}. For each object, given the assumption that the $a$ and $e$ values are consistent with being CCs, the free inclinations are also consistent with being CCs. It is interesting to note, however, that 2003 $\mathrm{ABCD}$ has a minimum semi-major axis of $a \approx 47.5$ au, suggesting that it may not truly be a CCKBO, but rather may be in a $2:1$ mean motion resonance with Neptune. If this is the case, the Hamiltonian model used to calculate its free inclination is incorrect, meaning that the result does not hold.
\begin{table}[ht]
\caption{Derived Orbital Limits}
\centering
\label{tab:object_parameters}
\renewcommand{\arraystretch}{1.3}
\setlength{\tabcolsep}{10pt}
\begin{tabular}{lccc}
\toprule
Object ID & 95\% confidence interval for $i_{\text{free}}$ assuming CC & $a_{\text{min}}$ (au) & $e_{\text{min}}$ \\
\hline
2003 $\mathrm{BF}_{91}$ & $1.08^{\circ}$--$1.62^{\circ}$ & $42.7$ & 0.014 \\
\hline
2003 $\mathrm{BH}_{91}$ &$0.09^{\circ}$--$1.01^{\circ}$ & $43.9$ & 0.031 \\
\hline
2003 $\mathrm{ABCD}$ & $0.49^{\circ}$--$0.53^{\circ}$ & $47.5$ & 0.026  \\
\hline
2003 $\mathrm{WXYZ}$ & $1.61^{\circ}$--$1.81^{\circ}$ & $45.3$ & 0.059 \\
\hline
\end{tabular}
\label{tab:free_inc}
\end{table}

\section{Discussion and Conclusions}
\label{sec:discussion}

\begin{figure}
    \centering
    \includegraphics[width=0.8\linewidth]{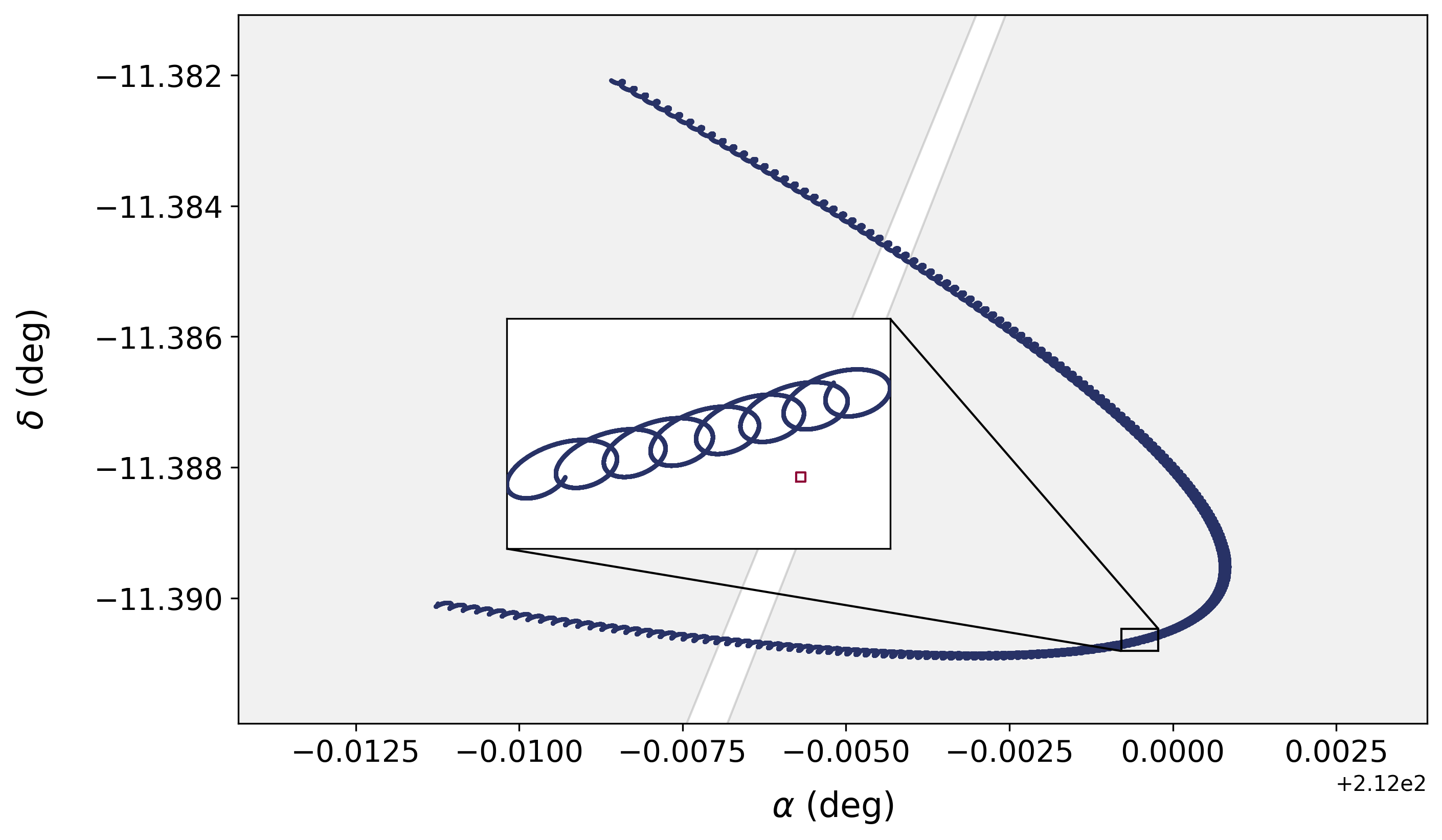}
    \caption{Trajectory of 2003 $\mathrm{BF}_{91}$ on the sky as seen by HST over the course of the survey. The inset plot emphasizes how nonlinear the trajectory is, due to HST's low Earth orbit; the proper motion in the sky plane remains linear for at most a few minutes at a time. The red square shows the size of one pixel; it is clear that the parallax loops are too large to simply ignore. The \texttt{heliostack} algorithm automatically corrects for this parallax. The grey regions represent the locations of different CCDs, showing that we are able to find objects that cross detectors over the course of the survey.}
    \label{fig:trajectory}
\end{figure}

In this paper we have introduced \texttt{heliostack} as a solution to the problem of nonlinear shift-and-stack, and applied it to archival HST data. We used the algorithm to exhaustively search for CCKBOs in a set of 1118 images taken over a 15-day time span, successfully recovering both of the known sub-threshold objects in the data, and adding two new discoveries. We did not perform a search for bright objects in these data, so we did not report measurements for 2000 $\mathrm{FV}_{53}$ and 2003 $\mathrm{BG}_{91}$, but we could see them in the images. As far as we are aware, 2003 $\mathrm{ABCD}$ and 2003 $\mathrm{WXYZ}$ are the first objects to be discovered in stacks of images with a time baseline longer than about one day. The algorithm did not have any problem finding objects that moved between chips, and was able to automatically correct for the extremely nonlinear on-sky trajectories of the objects caused by HST's orbits around Earth (see Figure \ref{fig:trajectory}).

The search process was extremely efficient; all of the steps discussed in this paper were run in a few hundred GPU hours on NVIDIA H200 cards, and a few thousand CPU hours. From start to finish, the process took less than one day on Harvard's Cannon cluster. Furthermore, in this case the production of a final catalog did not require any machine learning or visual inspection. A search of more than $10^{14}$ trajectories resulted in a list of just 145 candidates, where 141 of which were implanted sources, and other four are almost certainly real objects. This kind of purity may not be realistic when using datasets with variable image quality, but we expect that it should be fairly standard for images taken by high-quality space telescopes.

\texttt{heliostack} will be immensely helpful in elucidating science when applied to other datasets. In particular, it will enable deep searches for solar system objects in surveys with more data, and with sparser temporal cadences, than was previously possible. We plan to apply it to data from HST, JWST \citep{stansberry2021, trilling2025}, and eventually Roman. We will also apply it to select ground-based observations of sparsely-sampled deep drilling fields such as the Dark Energy Survey supernova fields \citep{des}, the DECam Ecliptic Exploration Project \citep{deep1}, and eventually LSST. It is also interesting to note that because of the way the algorithm is constructed, in which the data from different exposures are transformed to a common canvas, it is possible to combine heterogeneous datasets taken by different telescopes, enabling deeper searches than any one telescope can achieve.

In addition to being used as a general search tool, \texttt{heliostack} will be useful for several targeted applications in the discovery of solar system objects. One example is the discovery of faint targets for spacecraft such as New Horizons \citep{buie2024} and Lucy \citep{Levison2021, manzano2025}. Another application is doing targeted studies of specific orbital populations, for example CCKBOs (as in this paper), Neptune Trojans, Jupiter Trojans, Planet Nine, or anything else as long as the assumption of two-body motion holds. Since the number of required stacks can be calculated \textit{a priori} \citep{Holman2019}, the algorithm has a predictable computational demand, making it possible to decide whether any given study is feasible. Of course, there are still some limitations. For example, complete searches for certain dynamical classes, such as the NEOs, may prove to be computationally infeasible at the moment, but may become possible in the future. It is also not practical to stack images over arbitrarily long time spans; the number of required stacks grows roughly like the time span cubed \citep{bernstein2004, Holman2019}, so it is best to be discerning when deciding what data is worth stacking.

Finally, we note that while this paper has presented \texttt{heliostack} in the context of image data, there is no reason why it must be limited to pixels. Indeed, it is possible to treat both images and catalogs under this same framework; we can think of catalogs as being sparse images, or think of each pixel in an image as being an entry in a catalog. It is possible to use the \texttt{heliostack} framework to do trackletless linking in sparse datasets over long time baselines (such as DES and LSST) with well-defined computing requirements. We investigate the application of this technique in forthcoming work \citep{ruch2026}.

\begin{acknowledgments}
    We thank David Trilling for insightful feedback on this manuscript.
    
    K.J.N and M.J.H. gratefully acknowledge support from NSF grant No. AST2206194, NASA YORPD Program grant No. 80NSSC22K0239, and STScI grant JWST-GO-01568.004-A.

    This project is supported by Schmidt Sciences, LLC.
    
    The computations in this paper were run on the FASRC Cannon cluster supported by the FAS Division of Science Research Computing Group at Harvard University. 

    This research is based on observations made with the NASA/ESA Hubble Space Telescope obtained from the Space Telescope Science Institute, which is operated by the Association of Universities for Research in Astronomy, Inc., under NASA contract NAS 5–26555. These observations are associated with program GO-9433.

    Some/all of the data presented in this paper were obtained from the Mikulski Archive for Space Telescopes (MAST) at the Space Telescope Science Institute. The specific observations analyzed can be accessed via \dataset[https://doi.org/10.17909/axjs-nh20]{https://doi.org/10.17909/axjs-nh20}. STScI is operated by the Association of Universities for Research in Astronomy, Inc., under NASA contract NAS5–26555. Support to MAST for these data is provided by the NASA Office of Space Science via grant NAG5–7584 and by other grants and contracts.
\end{acknowledgments}

\bibliographystyle{aasjournal}
\bibliography{references}

\end{document}